\documentclass{article}
\usepackage{epsfig}
\usepackage{float}
\usepackage{bm,amsmath,amssymb,amsthm,latexsym,enumerate}
\usepackage[mathcal]{euscript}
\usepackage{hyperref}
\newcommand{\be}{\begin{equation}}
\newcommand{\ee}{\end{equation}}
\newtheorem{theorem}{Theorem}[section]
\newtheorem{proposition}[theorem]{Proposition}

\title{Blocking of 2D bistable reaction-diffusion fronts by obstacles}%

\author{
Jean-Guy Caputo \thanks{INSA Rouen Normandie, Normandie Universit\'e
LMI UR 3226, F-76000 Rouen, France E-mail: caputo@insa-rouen.fr},
Gustavo Cruz-Pacheco \thanks{Depto. Matem\'{a}ticas y Mec\'{a}nica, I.I.M.A.S.-U.N.A.M., Apdo. Postal 20--726, 01000 M\'{e}xico D.F., M\'{e}xico},
Jacek Gatlik \thanks{AGH University of Krakow, Faculty of Physics and Applied Computer Science,\\ 30-059 Krakow, Poland},
and
Benoit Sarels \thanks{Sorbonne Universit\'e, CNRS, Université Paris-Cit\'e,\\ Laboratoire Jacques-Louis Lions, 75005 Paris, France}
}
\begin{document}
\maketitle


\begin{abstract}
We investigate numerically the blocking of two-dimensional bistable 
reaction–diffusion fronts by geometric obstacles. Our goal is to 
derive quantitative criteria for front propagation in the presence 
of spatial heterogeneities. 
Using a conservation-law approach, we show that the integral of the reaction 
term acts as an effective driving force for the front. Combining this 
insight with the exact one-dimensional traveling wave solution, we construct 
a reduced analytical model that predicts blocking thresholds. 
In particular, we obtain explicit conditions for front propagation in 
a waveguide connected to a conical region of angle $\theta$, valid for 
widths $w \lesssim 4$. The model captures the influence of both 
geometry and nonlinearity, and shows good agreement with numerical 
simulations. 
Finally, we extend the analysis to more complex geometries, including 
checkerboard-like obstacles, and derive simple heuristic rules governing 
front propagation.
\end{abstract}

{\bf keywords} \\
bistable reaction-diffusion front ; obstacle ; reduced model

\tableofcontents

\section{Introduction}

Reaction diffusion equations are ubiquitous in chemistry and biology,
see the reviews by Scott \cite{Scott}, Murray \cite{Murray}, and the 
more recent review by Volpert \cite{Volpert}.
Examples are fungus propagation in crops of 
wheat or barley, the propagation of an electrical impulse  in a nerve 
and the propagation of an epidemic in a geographic network.
Many of these equations can only be solved numerically. In 1D however, two
important models have exact solutions:  the Zeldovich 
and the Fisher equations with nonlinearities that are cubic and 
quadratic respectively yielding bistable respectively monostable 
stationary states. These exact solutions provide valuable insight 
into the dynamics and 
serve as effective approximations of the system in the presence of 
perturbations.

A fundamental question concerns the interaction of such fronts with spatial 
heterogeneities.
Using a perturbation method together with extensive numerical calculations
the present authors analyzed \cite{cs11} the pinning of a kink 
by abrupt large amplitude spatially localized defects for the cubic 
bistable model. In \cite{ccs21} we analyzed stopping of a bistable kink by
a no reaction zone. On the other hand, a monostable front will 
always cross a non reaction region \cite{ccs21}. Another important defect
is an obstacle corresponding to an inaccessible region in the domain.
For example, a bistable reaction-diffusion front propagating in an 
inhomogeneous waveguide can
be blocked by a sudden enlargement, see Chapuisat and Grenier \cite{cg05} and 
Berestycki et al \cite{bbc16}. In physiology, the nerve impulse in a neuron
can be stopped by a sudden enlargement of the axon, see the nice discussion in
\cite{bbc16} and also \cite{thesebouhours}. See also the generalization
to periodic media done in \cite{dr18}.

The studies \cite{bbc16} and \cite{thesebouhours} prove passage or not 
passage depending on geometric conditions. However they do not give precise
values for the widths of the waveguide leading to blocking for a 
given nonlinearity. 
Therefore, the objective of the present study is to derive quantitative thresholds for front propagation in heterogeneous geometries. We address the 
following questions: \\
(i) What are the critical widths leading to blocking? \\
(ii) How do more complex obstacles affect propagation?\\
(iii) Can front blocking be predicted by a reduced analytical model ? 
\\
Using an approximate analysis based on the 1D front solution, we obtained
an analytical model describing the blocking of the front in the junction
between two waveguides of different widths $(w_1,w_2)$ or 
a waveguide connected to a cone of angle $\theta$.  This model
involves the integral of the reaction term over the front. Using it, we
can derive explicit values of the widths $(w_1,w_2)$ or $(w_1,\theta)$
causing blocking; this value also depends
on the parameter $a$ of the bistable nonlinearity. \\
The article is organized as follows. Section II presents the model,
the numerical method and numerical results showing that 
a straight waveguide connected to a cone can lead to front blocking.  
These numerical results are in good agreement with the analysis detailed
in section III. Section IV studies  two parallel waveguides connected to a cavity, as well as a checkerboard geometry. Conclusions are drawn in section V.

\section{The model and numerical results}

We consider the bistable reaction-diffusion model
\be\label{z2d}
u_t = \nabla ( b \nabla u )  + u(1-u)(a-u) ,\ee
in a 2D rectangular domain $\Omega = [0,L]\times [0,L] $  with 
homogeneous Neumann boundary conditions. The nonlinearity
$$ R(u) = u(1-u)(a-u),$$
is the standard bistable nonlinearity.
The term $b(x,y)$ is spatially dependent.
For $b=1$ in 1D \cite{cs11} we have the exact front solution
\be\label{front1D} u =  {1 \over 1+ \exp[\sqrt{1 \over 2}(x- ct)] }. \ee
The speed and width of the front are given by 
\be\label{speedWidth} c =  \sqrt{1 \over 2}(1-2a), ~~~ w = \sqrt{2}.  \ee
Note that $c=0$ for $a=0.5$.
In this study, we will consider specifically an inhomogeneous $b$. More
precisely the spatial dependance of $b(x,y)$ allows to represent 
different types of obstacles: $b=1$ outside the obstacle and 
$b << 1$ inside the obstacle.

We will examine two different numbers associated to the solution $u$:
the integral of $u$ and of the reaction term $R(u)$
\be\label{avi} 
<u> (t)\equiv {\int_\Omega u dx dy \over \int_\Omega dx dy } .\ee
\be\label{drive2}
\mathbf{R} \equiv \int_{\Omega} R(u) dx dy . \ee

\subsection{Radial solutions}

In 2D, an important class of solutions are radial fronts. In this section,
we show how they are related to the 1D exact solution (\ref{front1D}). 
Consider the static problem \eqref{z2d}, 
\be\label{sz2d}  
\Delta u   + R(u) = 0 ,
\ee
on a radial 2D domain of size $L$. 
Two numerical schemes were tested, a Newton-Raphson algorithm and the
relaxation scheme presented in \cite{bl1980}:
\be\label{sz2d}
-\Delta u^{k+1} +K u^{k+1}= K u^{k}  + R(u^k) ,
 \ee
where $K>0$ is a constant.
We found that imposing 
$u_r = 0$ at $r=0$ always leads to a zero solution. This is consistent with 
the fact that a radial front will always propagate so that there are no static
radial fronts.

\begin{figure}[H]
\centerline{ \epsfig{file=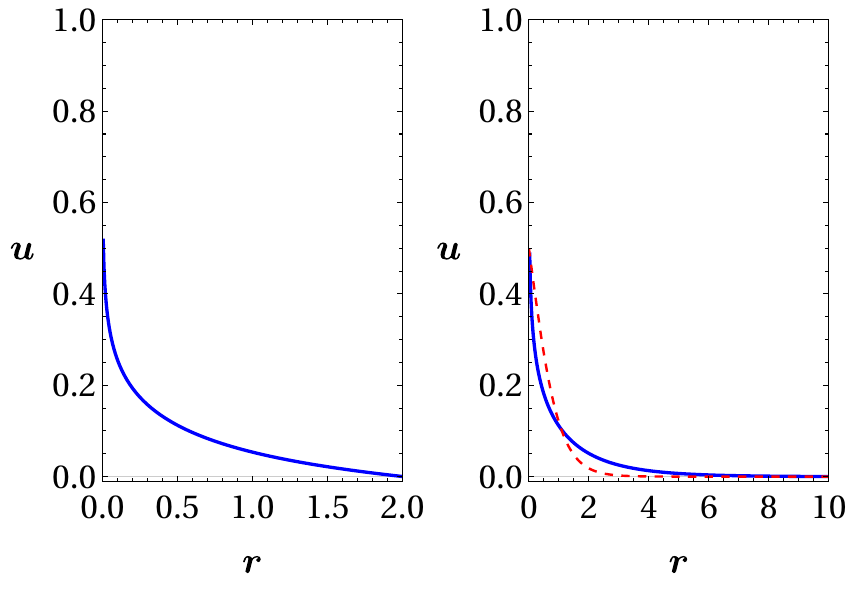, height = 5 cm, width=8 cm, angle=0} }
\caption{Solution of the static problem for $L=2$ (left) and $L=10$ (right).
The parameter $a=0.3$.}
\label{ur}
\end{figure}

To obtain a nonzero static solution, we need to fix $u=u_0$ at $r=0$. Results 
are shown in Fig. \ref{ur} for $L=2$ (left panel) and $L=10$ (right panel). We 
chose $n=400$ grid points and used the finite difference 
scheme of Strickwerda \cite{sn86} with $K=2$.
The scheme converges rapidly, reaching a Cauchy residual of $10^{-9}$
in about 100 iterations. 
For this type of nonlinearity, there is no minimal radius to obtain a solution
\cite{bl1980}, so there is always a solution for any $L >0$.
Clearly for $L=2$, the domain is too narrow to "fit" the solution, we need to
increase it so that the solution and its tangent are close to zero for
$r=L$.
In the right panel of Fig. \ref{ur}, we included the
kink profile centered on $x_0=0$ and with a width $w_0=0.5$. Note the
good agreement with the static solution. \\
The typical scale of variation of the 2D solution is about 5, we then
expect 2D effects to appear for inhomogeneities larger than that value.
This will be confirmed in the following sections.

\subsection{Numerical method}

We study numerically the case of a waveguide connected
to a cone of angle $\theta$. In particular, we explore the parameter
space $(w, \theta)$. The results will be compared with the analytical 
predictions of the next section.
We integrate the partial differential equation \eqref{z2d}
using the method of lines where the spatial part is a finite volume
discretisation and the time advance is done using an ordinary differential
equation solver, typically a fourth order Runge-Kutta method. On a square grid of size $L$
we consider $u_{i,j} \equiv u(i dx, j dy)$. Integrating the operator over
a cell of area $dx \times dy$ centered on $(i,j)$ yields
$${\dot u_{i,j}} = 
{1\over dx^2} [(u_{i+1,j}-u_{i,j})b_{i+1/2,j} - (u_{i,j}-u_{i-1,j})b_{i-1/2,j}]+$$
\be\label{disz2d} 
{1\over dy^2} [(u_{i,j+1}-u_{i,j})b_{i,j+1/2} - (u_{i,j}-u_{i,j-1})b_{i,j-1/2}]+
R(u_{i,j}),\ee
where $b_{i-1/2,j} \equiv {1 \over 2}(b_{i-1,j}+b_{i,j})$ and similarly for
the other $b$ terms.
The resulting system is then integrated using a fourth-order Runge-Kutta scheme.
There is a Courant-Friedrich-Lewy stability condition which reads
$$ {dt \over dx^2} < {1\over 2}.$$

In practice, we chose $L=100, ~dx=dy=0.1$ and $dt=10^{-3}$. 
The values of $b$ are 
$b=1$ in the accessible region and $b \ll 1$ inside obstacles, effectively 
enforcing no-flux boundaries. 
Unless otherwise stated, we fix the bistable 
parameter to $a=0.3$, corresponding to a right-propagating front.
A run for a duration $T=400$ takes about 3 hours on an 2.7 GHz Intel 
processor. The code has been
ported to a graphics card (GPU) and it runs in about 5 minutes showing
a considerable acceleration. This speedup is likely due to the 
simplicity of the algorithm.

\subsection{Waveguide connected to a cone}

\begin{figure}[H]
\includegraphics[width=\linewidth]{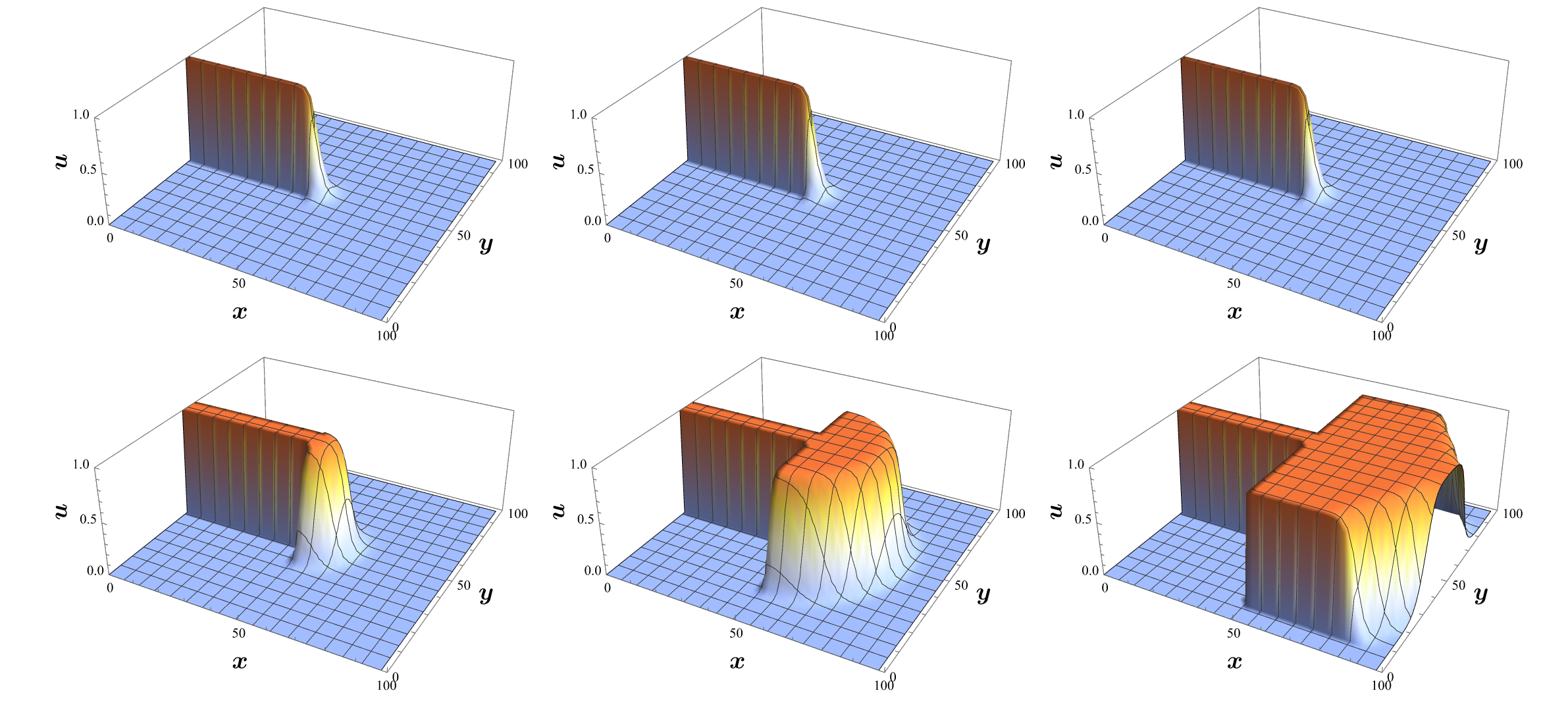}
\caption{Propagation of a kink in a T waveguide: three snapshots at times
$t=160, 240$ and $320$ for $w_1=4$ (top panels) and $w_1=10$ (bottom panels).
}
\label{w2a10a40}
\end{figure}

We assume that the waveguide on the left is connected
to a cone of angle $\theta$ as shown in Fig. \ref{wavguid3}.
We first consider the geometry $\theta = \pi/2$, i.e.
a waveguide of left width $w_1$ and right
width $w_2$ with a sharp transition at $x=0$.
Fig. \ref{w2a10a40} shows three snapshots at instants $t=160,~240$ and $320$ 
of $u(x,y,t)$ for a waveguide $w_1=4$ (top panels) and 
$w_1=10$ (bottom panels). The width of the waveguide on the right is $w_2=80$.
The top panels ($w_1=4$) show that the front becomes blocked at the transition.
For the wider waveguide shown in the bottom panels, the front crosses
into the wider region and continues its progression. 
\begin{figure}[H]
\includegraphics[width=\linewidth]{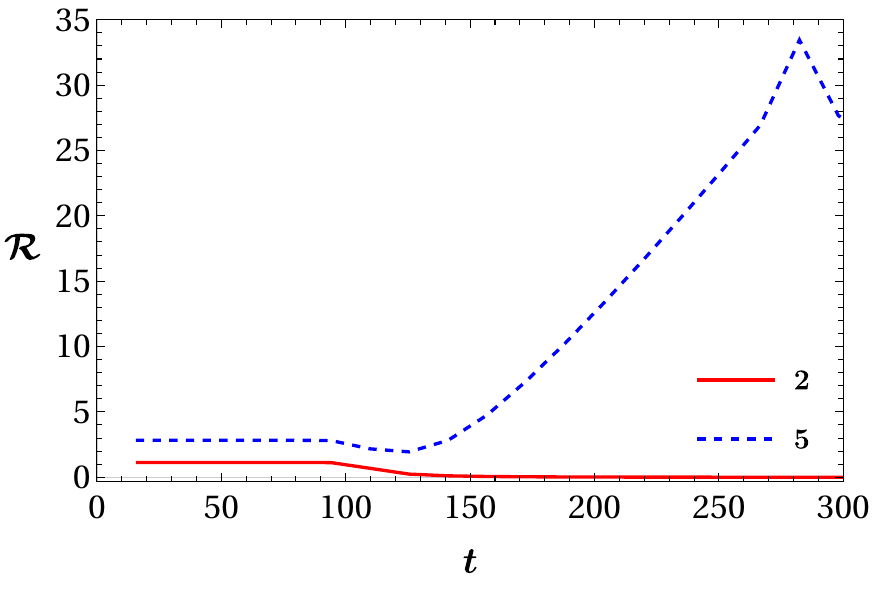}
\caption{Time evolution of the integral of the reaction term $\mathbf{R}$
from (\ref{drive}) over 
the computational domain for the solutions shown in Fig. \ref{w2a10a40}.
}
\label{int25a40}
\end{figure}
Fig. \ref{int25a40} shows the integral of the reaction term $\mathbf{R}$
on the computational domain from (\ref{drive}).
This integral converges to 0 for $w_1=4$ and increases sharply for
$w_1=10$.

\begin{figure}[H]
\includegraphics[width=\linewidth]{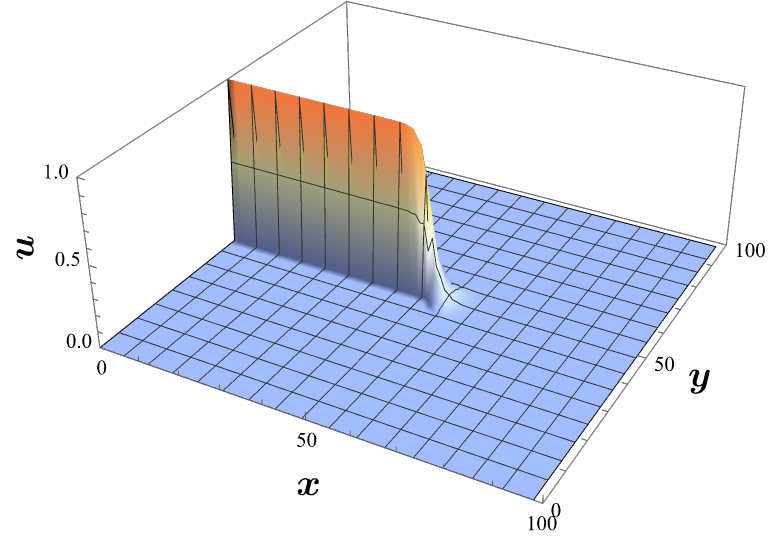}
\caption{Blocking of a front by a cone of angle $\theta =0.75$.
The width of the waveguide is $w=2$.}
\label{w1teta075}
\end{figure}

Consider now that the angle $\theta$ can be varied between 0 and $\pi$.
For simplicity we write $w=w_1$.
Kinks can get blocked for waveguides with angles smaller than $\pi/2$. 
Such a blocking is shown in Fig. \ref{w1teta075} for $\theta = 0.75$
and $w_1=2$. Notice how the front becomes blocked at the
interface.
\begin{figure}[H]
\includegraphics[width=\linewidth]{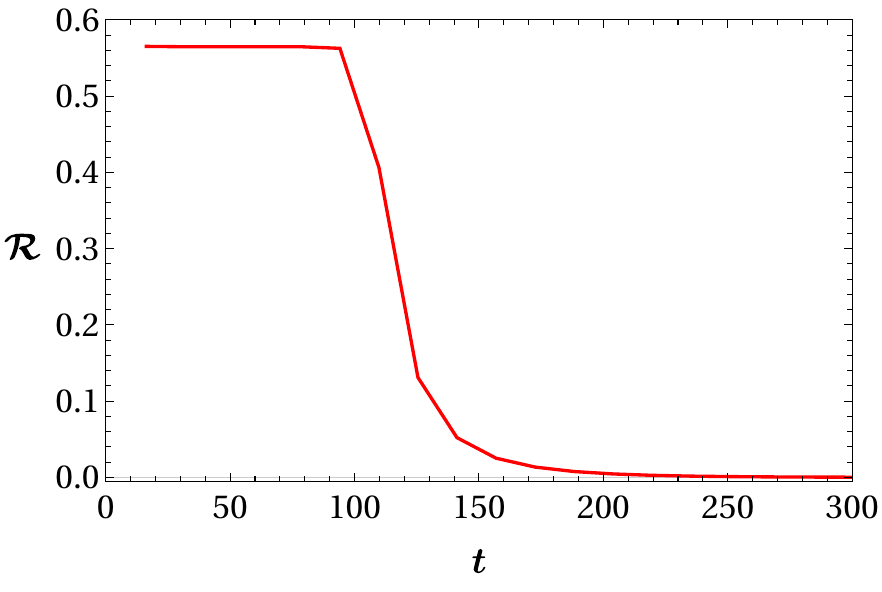}
\caption{Time evolution of the integral of the reaction term $\mathbf{R}$
for the solutions shown in Fig. \ref{w1teta075}.}
\label{int075}
\end{figure}
Fig. \ref{int075} shows the integral of the reaction term $\mathbf{R}$
for the numerical solution shown in Fig. \ref{w1teta075}.
This integral converges to 0 confirming the blocking of the front.

\begin{figure}[H]
\includegraphics[width=\linewidth]{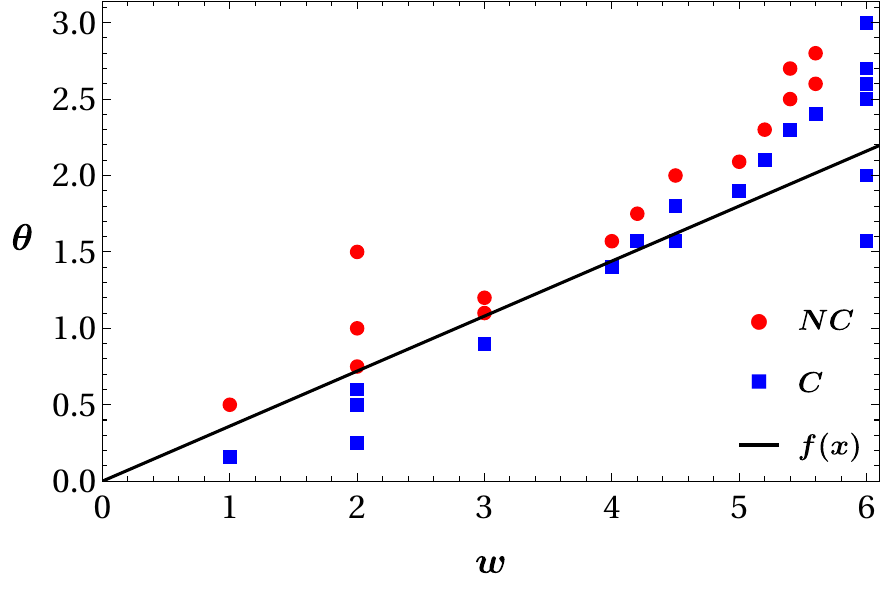}
\caption{Parameter plane $(w_1,\theta)$ showing crossing (blue squares) 
vs blocking of the front (red $\times$).}  
\label{pnp}
\end{figure}

The parameter plane $(w,\theta)$ has been explored systematically
leading to Fig. \ref{pnp}. The couples $(w,\theta)$ for which the
kink crosses over are indicated as squares (blue online) and the ones for
which the kink gets blocked are shown as $\times$ (red online). 
The analysis of the next section can be used to explain these results.
Using equation (\ref{IsIr}) from section 3.3 and plugging in the parameter $a$, it follows that
\be \label{IsIra} 
r_\theta = w 0.2 \sqrt{2}+ 2 \theta[ -0.5 +0.16 \log(2)]\approx 0.28 w -0.78 \theta. \ee  
The condition $r_\theta =0$ yields the linear threshold
\be \label{IsIra2} \theta \approx 0.36 w . \ee
This line is plotted in black in Fig. \ref{pnp}. One can see the good
agreement of this expression with the threshold for the blocking of the kinks
for $w < 5$.

For $w>5$ the kink always crosses into the cone
and expression (\ref{IsIra2}) does not seem to apply. For such large $w$,
the curvature of the front in the cone cannot be neglected. This is typical
of the propagation of nonlinear waves in a waveguide of cross-section $w$,
see for example \cite{cfgv96} where we considered how 2D corrections
affect sine-Gordon kinks for Josephson junctions. We give more details in
the next subsection.

\section{Analysis}

In a recent article Berestycki, Bouhours and Chapuisat \cite{bbc16} proved that a bistable front will always propagate in a waveguide with decreasing cross section. By contrast, if the waveguide opens up abruptly, they proved that there can be blocking. They consider "blocked" solutions and show that these can exist in some configurations.  In this article, we use a direct argument based on conservation laws 
derived from the partial differential equation to distinguish between blocking and propagation of the front.

\subsection{Two-dimensional effects in a waveguide}

We consider a waveguide of cross-section $w$ extending from
$y=-w/2$ to $y=w/2$ and the equation 
\be\label{z2db}
u_t =\Delta u + R(u)  .\ee
Following \cite{cfgv96}, we decompose $u$ as 
$$u = U_0(x,t) + \phi(x,y,t),$$
where $U_0$ is such that
\be\label{z1d}
{U_0}_t ={U_0}_{xx} + R(U_0)  ,\ee
and $\phi$ is small.

Linearizing \label{z2db} around $U_0$, one obtains the following equation for
$\phi$
$$ \phi_t = \Delta \phi + dR(U_0)  \phi .$$
One can then expand 
$$ \phi = \phi_1(x,t) \cos ({\pi y \over w}) + \phi_2(x,t) \cos ({2\pi y \over w}) + \dots$$
and get the evolution of the amplitudes $\phi_1, \phi_2, \dots$ as
$${\phi_1}_t = {\phi_1}_{xx} -\phi_1 ({\pi \over w})^2 + dR(U_0) \phi_1,$$
$${\phi_2}_t = {\phi_2}_{xx} -\phi_2 ({2\pi \over w})^2 + dR(U_0) \phi_2 .$$
The term $dR(U_0)$ is $x$ dependent and acts as a potential exciting 
the amplitudes $\phi_1, \phi_2, \dots$. Neglecting it , one can use
a Fourier expansion to find the evolution of $\phi_1$. 
This yields
$$\phi_1 \sim e^{ikx -\omega_1 t},$$
where $$\omega_1 = k^2 + ({\pi \over w})^2 .$$
This shows that the smaller $w$ is, the faster is the decay of $\phi_1, \phi_2, \dots$. For a given $w$, we need $n = w/(2 \pi)$ modes $\phi$ to describe
correctly the 2D solution. As a consequence, the typical width $w$ above 
which one expects 2D effects is
$$w^* = 2 \pi.$$
A more refined analysis would give effects that depend on the nonlinearity
but this is a correction. The main effect comes from the geometry. \\
In view of this, it is not surprising that for $w \geq  w^*$ fronts
will cross into the cone, no matter the angle $\theta$.

\subsection{Waveguide with sharp transition}

\begin{figure}[H]
\includegraphics[width=\linewidth]{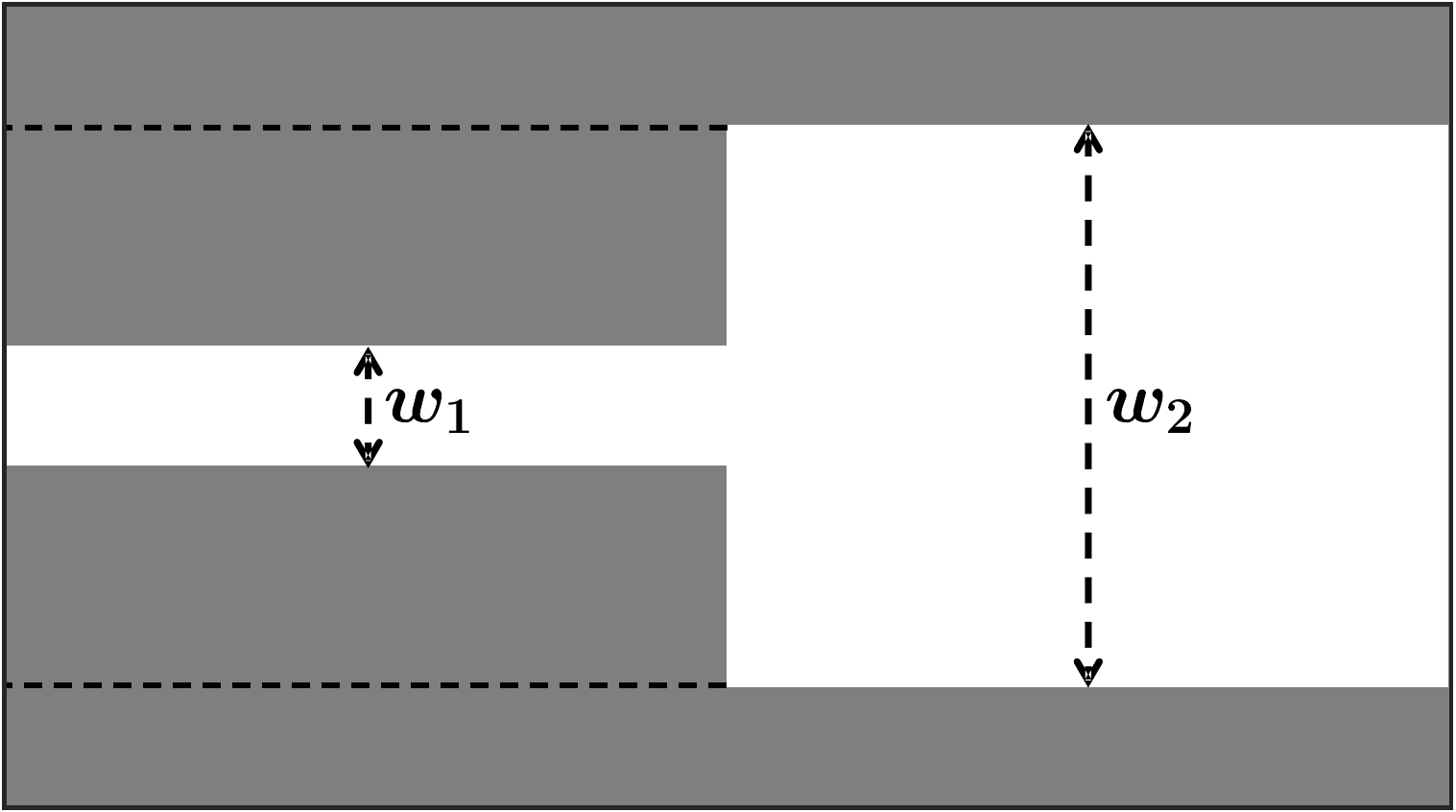}
\caption{An inhomogeneous waveguide: the gray area corresponds to $b<<1$ fixing
the boundaries of the waveguide.}
\label{wavguid}
\end{figure}

\begin{figure}[H]
\includegraphics[width=\linewidth]{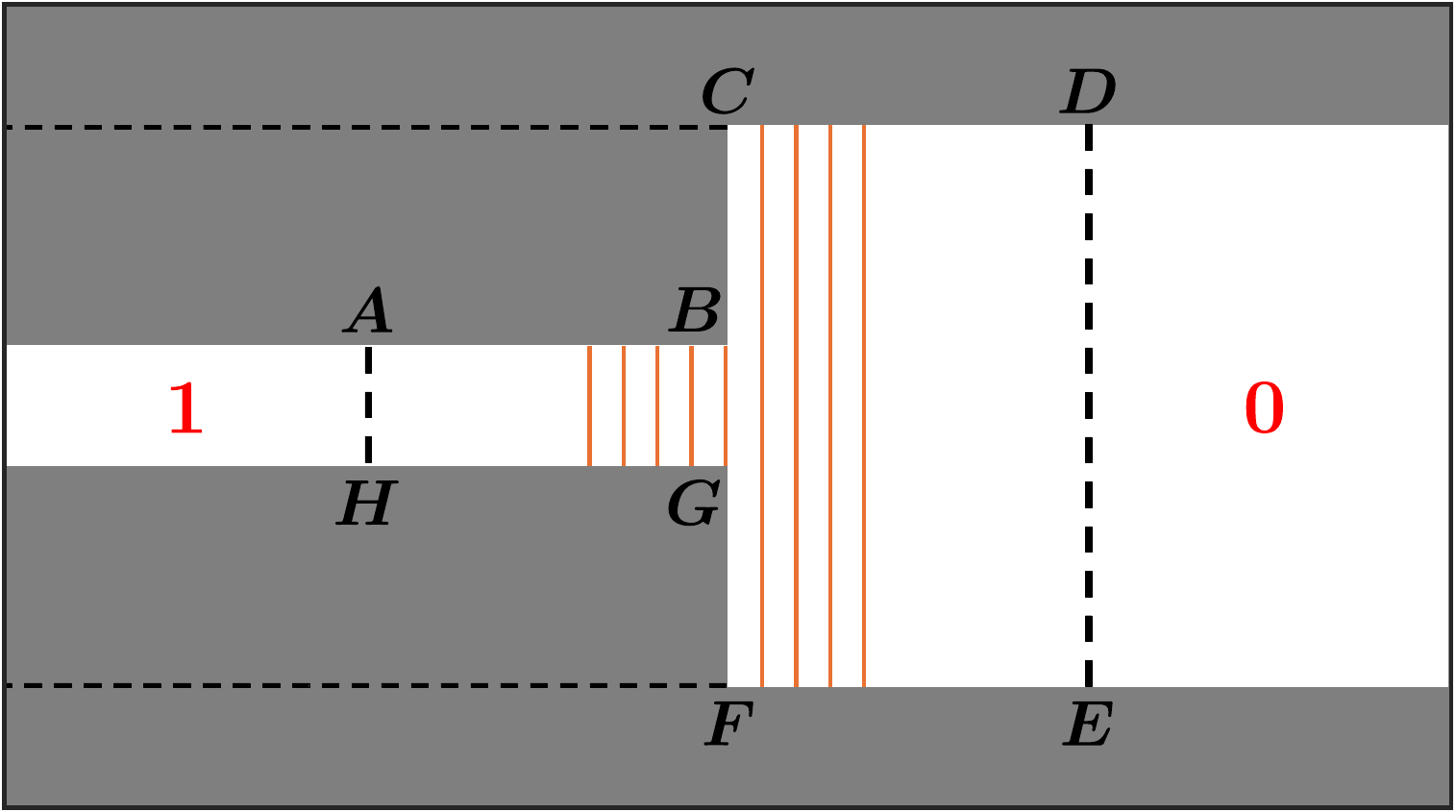}
\caption{Analysis of a trapped kink in a waveguide: the contour lines
are shown in red.}
\label{wavguid2}
\end{figure}

To simplify the discussion, we start with a junction between
a left waveguide of cross-section $w_1$ and a right waveguide of
cross-section $w_2$ as shown in  Fig. \ref{wavguid}.
The 2D reaction-diffusion equation (\ref{z2d}) can be written as 
\be\label{z2da}
u_t =\Delta u  + R(u),\ee
where the domain $\omega=ABCDEFGH \subset \Omega$ is shown in Fig. \ref{wavguid2}. 
To analyze the motion, we derive a conservation law over $\omega$.
For this, we integrate the left hand side of the equation on this domain and
obtain
\be\label{int}  \partial_t (\int_{\omega}u dx dy) = 
\int_{AH} \nabla u \cdot n ds
+\int_{DE} \nabla u \cdot n ds
+ \int_{\omega}R(u) dx dy, \ee
where we used the homogeneous Neumann boundary conditions to eliminate
the contributions of the boundaries $ABCD$ and $AGFE$.

Assume a front type solution, then $u \to 1, ~~\nabla u \to 0$ on the left and
$u \to 0, ~~\nabla u \to 0$ on the right. Then, the two flux integrals in the
right hand side of equation \eqref{int} are zero. We then obtain the following. 
\begin{proposition}
Let $u(x,y,t)$ be a sufficiently smooth solution of the 
reaction–diffusion equation 
$$u_t = \Delta u + R(u)$$
in a domain $\omega$, satisfying 
homogeneous Neumann boundary conditions. Assume that $u$ has a 
front-like structure with asymptotic states 
$u \to 1$ on the left and $u \to 0$ on the right. Then
\[
\frac{d}{dt} \left ( \int_\omega u , dxdy \right ) = \int_\omega R(u), dxdy.
\]
\end{proposition}
This result has the following consequences.
\begin{enumerate}
\item First, a front will move in the transition
region $\omega$ if $ \partial_t (\int_{\omega}u dx dy)>0 $ i.e., if  
$\mathbf{R}\equiv \int_{\omega}R(u) dx dy >0$. 
We interpret $\mathbf{R}$ as an effective driving force 
governing front propagation.
\be\label{drive}
\mathbf{R} \equiv \int_{\omega} R(u) dx dy . \ee
\item If a front moving inside $\omega$ is such that
$\mathbf{R}=0$ at some instant, then it will stop.
\item This argument can be made general, in particular
we can describe complex transition regions.
\end{enumerate}

To obtain quantitative results on the simple geometry we considered,
one should evaluate the right hand side of equation (\ref{int}). 
To this end, we introduce an approximate form of the solution.
The simplest assumption is that the kink has no $y$ dependence
so that we can use the exact solution and the 
1D analysis developed in our work \cite{cs11}.
We assume that the kink follows the 1D exact solution 
given by (\ref{front1D})
centered on an $x$ position $h$. Then equation (\ref{int}) reduces to 
\be\label{int2}
\partial_t (\int_{\omega}u dx dy) =
w_1 \int_{-\infty}^h R(U) dx + 
w_2 \int_h^{+\infty} R(U) dx \equiv r(h).\ee
Using the results of \cite{cs11}, we obtain
\be\label{int3}
r(h)= w_1 ( {1\over 2} -a)
 + (w_1-w_2) \left ( {a \over 1+e^h}
- {1 \over 2(1+e^h)^2} \right ). \ee
If there exists a position $h$ such that $r(h)=0$ then the kink
can get trapped inside $\omega$.

If we note 
$$ s = {1 \over 1 + e^h} , $$
then the part depending on $h$ reads
$$a s -{1 \over 2} s^2 ,$$
whose maximum is $a^2/2$ for $s=a$. Hence, one can derive the condition
for front pinning across a transition region region $w_1$ to $w_2$
\be\label{cond1}
w_2 \ge w_1 (1 + {1-2 a \over a^2}) . \ee

\begin{figure}[H]
\includegraphics[width=\linewidth]{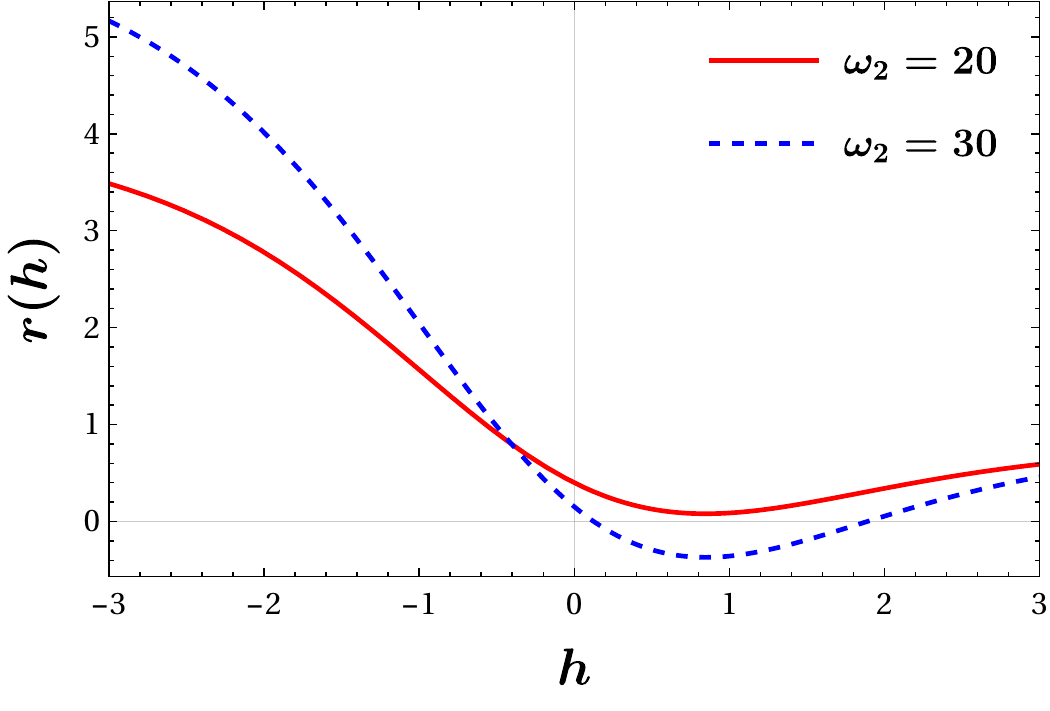}
\caption{Plot of $r(h)$ for $w_1=4, ~w_2=20$ and $w_2=30$.}
\label{rh}
\end{figure}

Fig. \ref{rh} shows a plot of $r(h)$ for $w_1=4$ and for 
two values of $w_2$, namely 20 and 30.
We clearly see that $r(h)$ for $w_2=30$ goes through
zero so that the kink gets trapped. 
Formula \eqref{cond1} gives the following: for $w_1=4$ and $a=0.3$ we 
obtain $w_2 > 21.8$  which explains that a front will cross for 
$w_2=20$ and get pinned for $w_2=30$.
Note that if the nonlinearity is multiplied by a factor $s$ then the equation
can be rescaled using the transformation $x'= x \sqrt{s}, ~y'=y \sqrt{s}$.
This means that the threshold value for crossing is now $w'= w/\sqrt{s}$, in
other words if the nonlinearity is multiplied by 4, the crossing threshold
is divided by 2. This scaling was also noted by the authors of \cite{bbc16}.

\subsection{Waveguide connected to a cone }

\begin{figure}[H]
\begin{center}
    \includegraphics[width=0.75\linewidth]{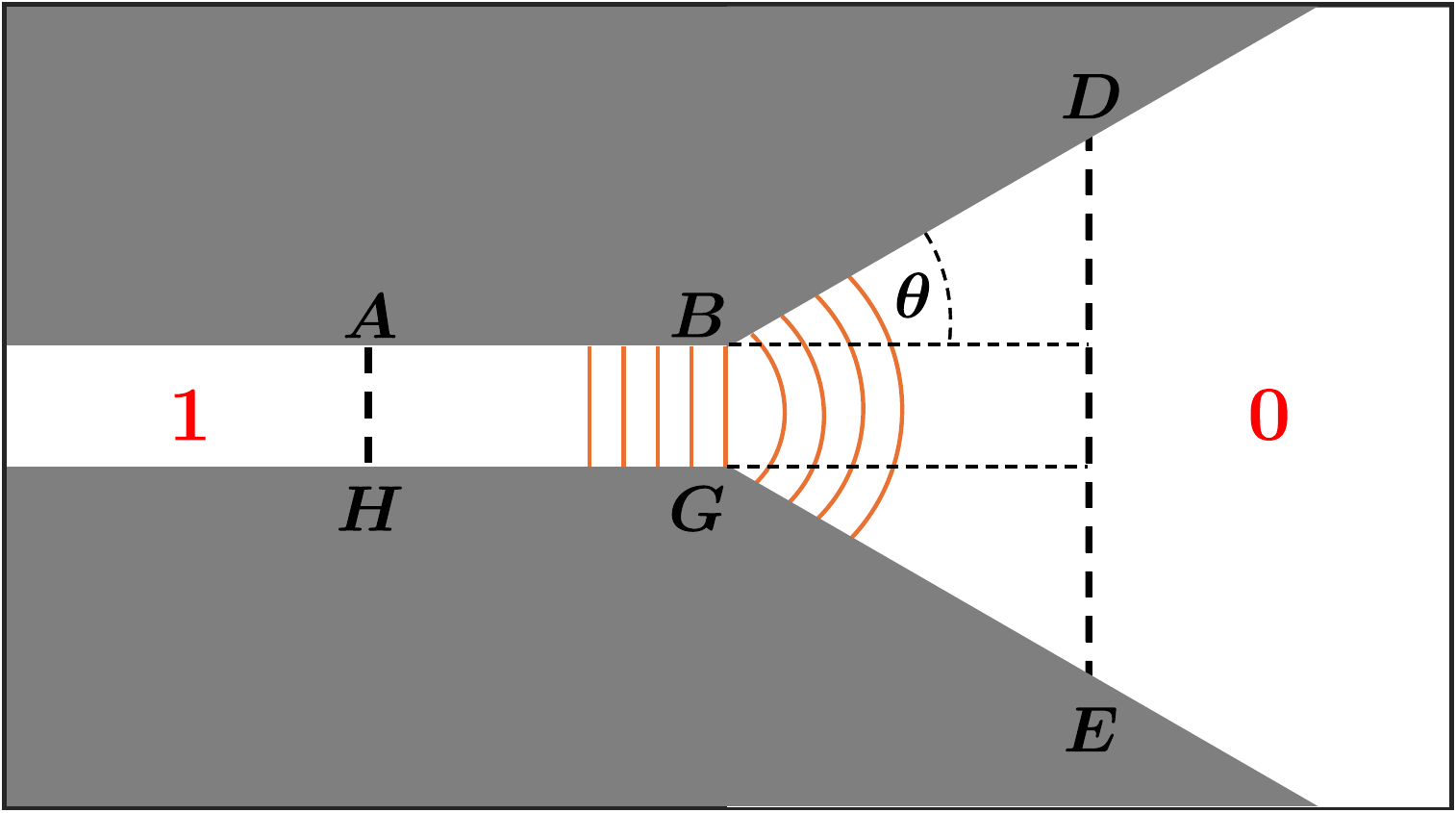}
\end{center}
\caption{Analysis of a trapped kink in a waveguide connected to
a cone: the contour lines are shown in red}
\label{wavguid3}
\end{figure}

We now generalize the previous geometry by replacing the rectangular right-hand region with a cone, as shown in Fig. \ref{wavguid3}. 
The analysis is more involved than the one for the straight waveguides
because of the 2D effects. In Fig. \ref{wavguid3} the rectangular region
ABGH can be extended by the rectangle in dashed lines on the right hand
side. This gives rise to one component of the integral. The bottom
and top triangular regions BCD and GEF will contribute to another component.
Starting from \eqref{int2} we obtain
\be\label{int4}
\begin{gathered}
    r_\theta \equiv \partial_t (\int_{\omega}u dx dy) = I_s + I_r =\\ 
w \int_{-\infty}^{+\infty} R(U) dx +
2 \theta \int_0^{+\infty} R(U) r dr .
\end{gathered}
\ee
Using the change of variable $z={x-x_0 \over \sqrt{2}}$ the first term in 
the right hand side can be calculated as 
$$ I_s = \sqrt{2} w \int_{-\infty}^{+\infty}  R(U(z)) dz = 
w\sqrt{2} ( {1 \over 2}-a) ,$$
where we used the results of \cite{cs11}.


Using the change of variable
$z={r-x_0 \over \sqrt{2}}$ in the integral $I_r$ of (\ref{int4}), one obtains
$$I_r = 2 \theta \int_{-x_0 \over \sqrt{2}}^{+\infty} 
(z \sqrt{2} + x_0) \sqrt{2} R(U(z)) dz .$$
To simplify the analysis, we set $x_0=0$ in the above integral so that
it becomes
$$I_r \approx 4 \theta \int_0^{+\infty} z R(U(z)) dz .$$
Using the expressions given in the appendix, we obtain the final expression
\be \label{IsIr} 
r_\theta \equiv I_s+I_r = w\sqrt{2} ( {1 \over 2}-a)  + 4 \theta  [ -{1 \over 2} -(2a -1) \log(2)  ].\ee
This parameter $r_\theta $ indicates whether the kink will cross into the cone
or not. If $r_\theta \leq 0$, there is no crossing.

At this point several remarks can be made:
\begin{itemize} 
\item A key quantity in estimating $r_\theta$ is 
$I_s/ w=\sqrt{2} ( {1 \over 2}-a)$ which is positive. The second term is 
negative for $a<0.5$.
We then expect to have crossing for small $w$ if $I_s/ w$ is large and vice
versa. This term $I_s/ w$ is the integral of the reaction term for a kink
in the 1D case. 
\item Formula \ref{IsIr} shows that the parameter $a$ of the nonlinearity
plays an important role. If $a \to 1/2$ the first term tends to zero and the
second reduces to $-\theta$ so that there is crossing only for very large $w$.
For $a=0.3$ we observed crossing for $w=4$ and $\theta =1.4$ whereas there is
no crossing for $a=0.4$.
\item The approximation $h \approx 0$ to compute the right hand side of the
integral $I_r$ is reasonable. See for example, Fig. \ref{rh} which shows that
$r_h \approx 0$ for $h \approx 0$.
\item Expression \eqref{IsIr} obtained in 2D can be generalized to 
arbitrarily large dimensions $n$. From expression \eqref{int2}, we 
expect the following scalings to hold
\be \label{IsIrn} 
r_\theta^n \equiv I_s+I_r = w^{n-1}\sqrt{2} ( {1 \over 2}-a)  + w^{n-2} 4 \theta  [ -{1 \over 2} -(2a -1) \log(2)  ].\ee
This indicates that the width for which the kink will cross should be 
independent of the dimension. 
\end{itemize} 

\section{Other Obstacles}

In this section, building on the numerical results presented above
and the analysis, we consider more complicated obstacles. Before doing
that, we analyze how a hole affects the front propagation.

\subsection{A single hole}

{\bf Effect of the radius} \\

\begin{figure}[H]
\centerline{ \epsfig{file=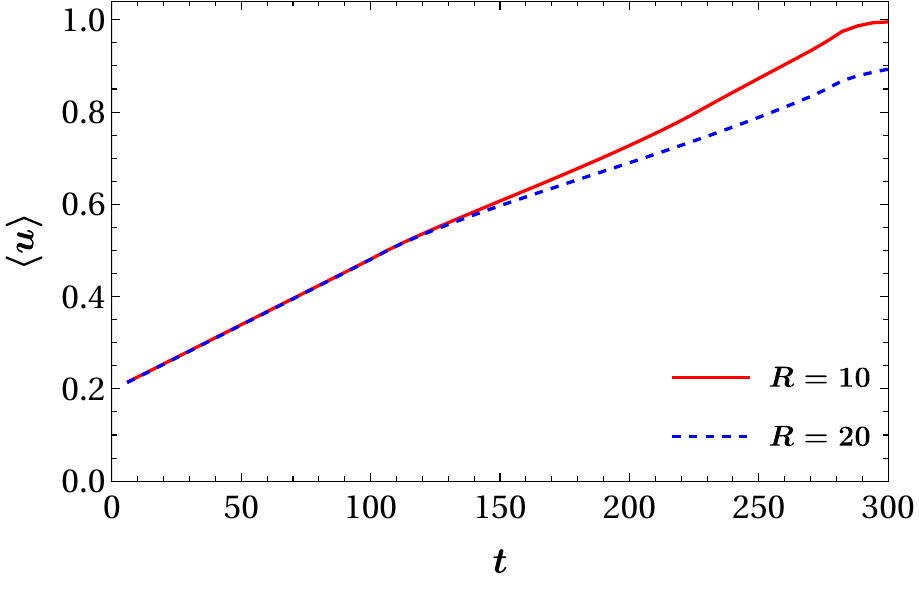, width=\linewidth} }
\caption{Plots of ${<u>(t)}$ for $R=10$ and 20.}
\label{r10r20}
\end{figure}

As expected a larger hole will slow down the front more than
a smaller hole. Fig. \ref{r10r20} shows ${<u>(t)}$ for $R=10$ and 20:
as expected the integral grows slower for $R=20$.

\subsection{Two waveguides in parallel}

\begin{figure}[H]
\centerline{ \epsfig{file=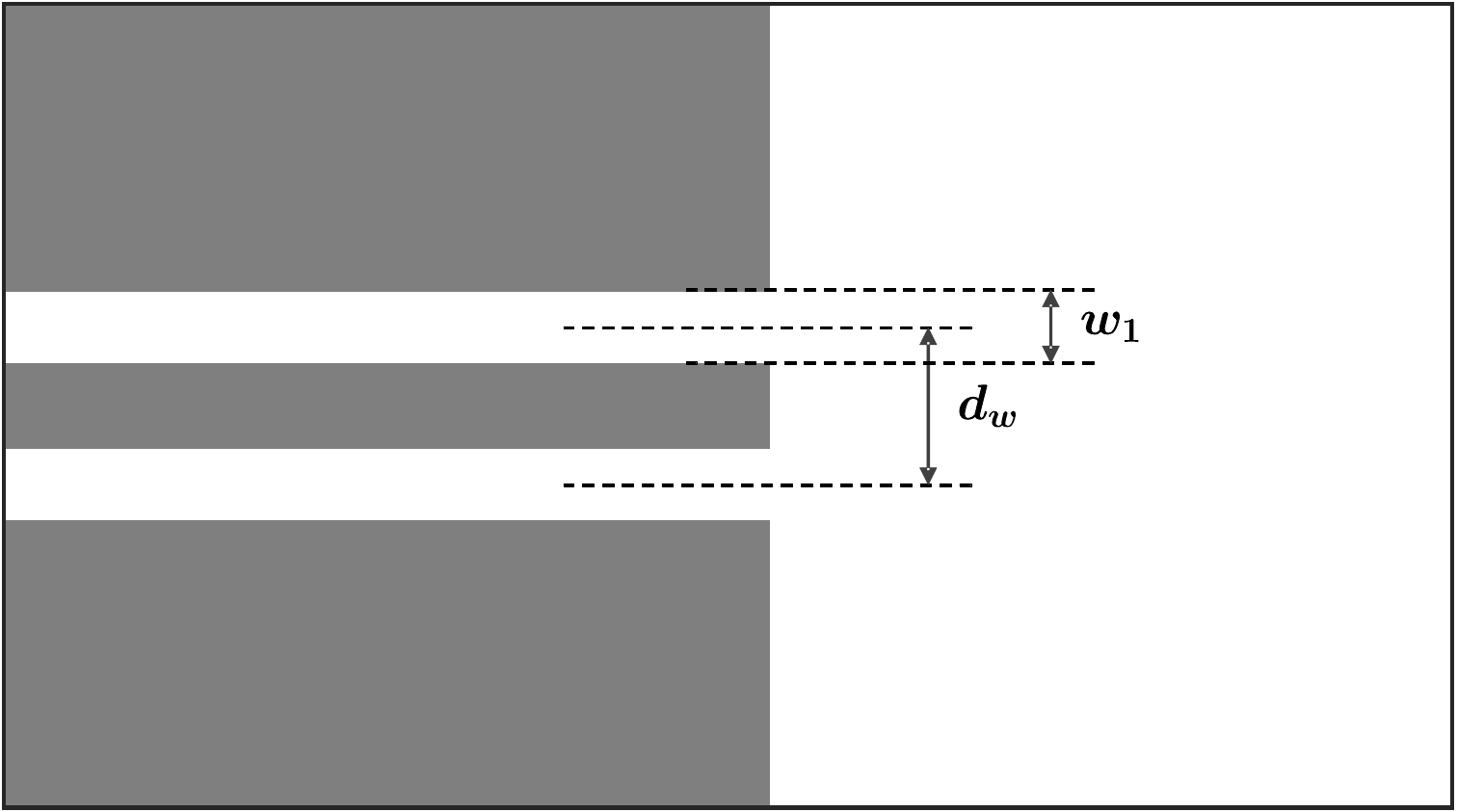, width=0.75\linewidth} }
\caption{Two waveguides of width $w_1$ connected to a large cavity}
\label{twowavg}
\end{figure}

We now examine how two fronts in parallel waveguides interact when
the waveguides connect to a large cavity. The geometry is shown in
Fig. \ref{twowavg}. 
Depending on the separation $d$ between the two waveguides, the fronts 
may either enter the cavity or become blocked.

\begin{figure}[H]
\includegraphics[width=\linewidth]{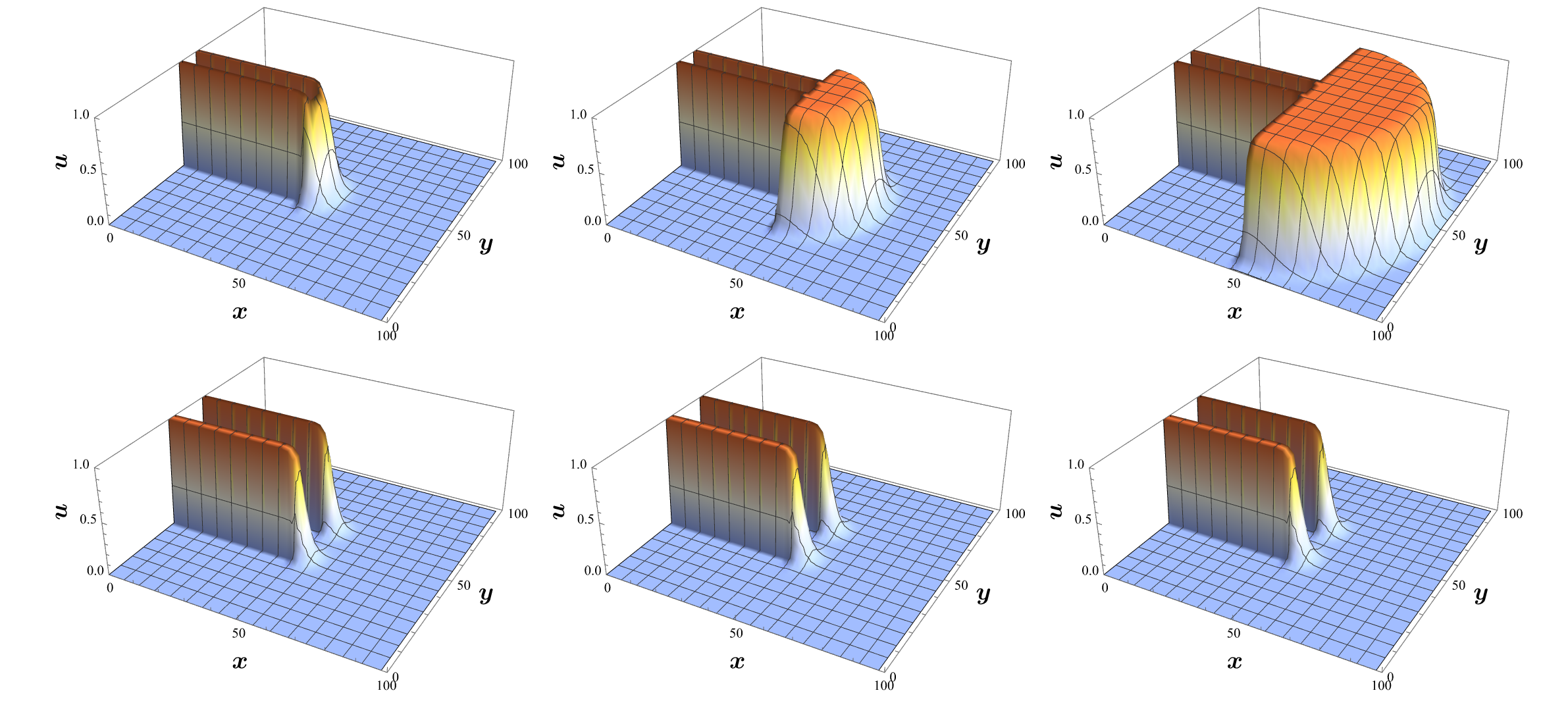}
\caption{Kink evolution in two waveguides of width $w=4$ separated by 
$d = 5$ (top panels) and $d = 10$ (bottom panels).}
\label{ntwowavg}
\end{figure}

Fig. \ref{ntwowavg} shows snapshots of the numerical solution for times
$t= 10,~15$ and 20 for $d = 5$ (top panels) and $d = 10$ (bottom panels). The
waveguides have width $w=4$ so a single such waveguide leads to blocking of the
front. The remarkable effect is that the blocking disappears and 
the two kinks cross into the cavity for $d = 5$. For $d = 10$, the kinks
are trapped.

\begin{figure}[H]
\includegraphics[width=\linewidth]{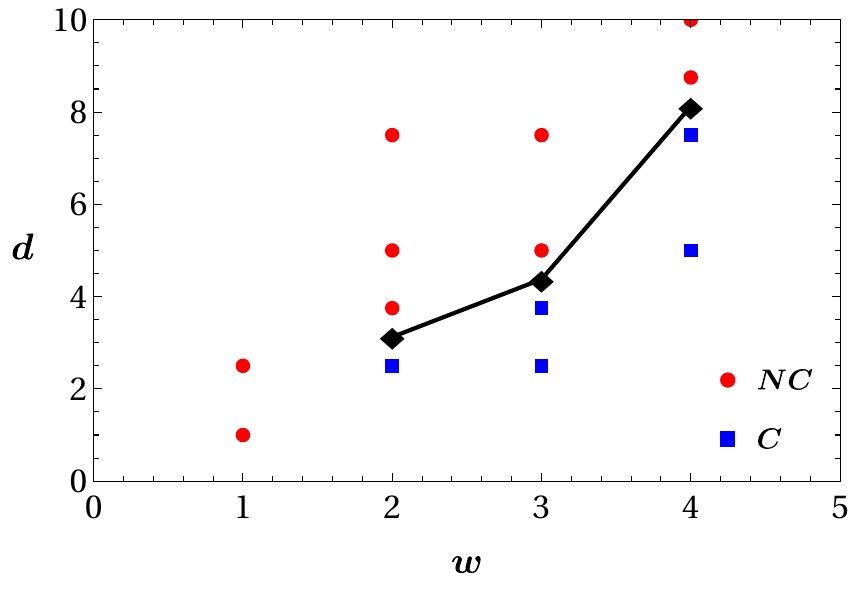}
\caption{Parameter space $(w,d)$ showing crossing (C) vs no crossing (NC)
for two kinks in waveguides of width $w$ separated by $d$.}
\label{wdpnp}
\end{figure}

We conducted a detailed exploration of the parameter space $(w,d)$ and
found that for $w>4$ the kinks always cross into the large area, as expected
from the results on a single waveguide.
For $w<4$, however the crossing of the two kinks depends on $d$. For
large $d$, there is no crossing while for smaller $d$ we observe crossing.
The results are reported in Fig. \ref{wdpnp}, where we place a square for
crossing (C) and a cross for no crossing (NC). The black line is drawn to
guide the eye. For $w>4$ we expect always to have crossing because the single
waveguide leading into a cone gives crossing for $\theta =\pi/2$.

\subsection{Series of holes: checkerboard}

\begin{figure}[H]
\includegraphics[width=\linewidth]{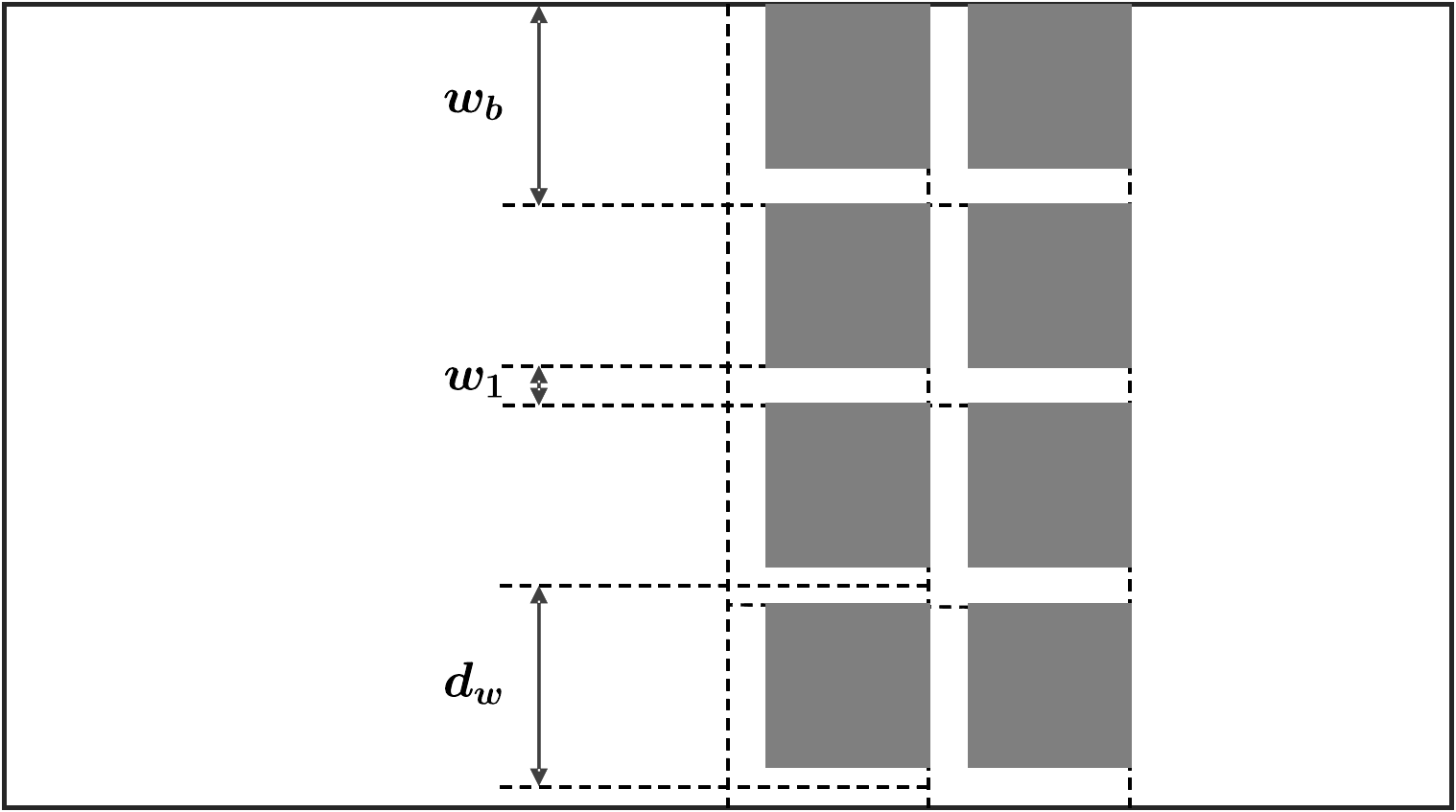}
\caption{Schematic drawing of a checkerboard obstacle.}
\label{check}
\end{figure}

We now consider a checkerboard-like obstacle constructed from several 
square blocks.
A typical configuration is shown in Fig. \ref{check}. The defect
is localized in a region $ x_{max}/2 \le x \le w_b$. It is such that
$b=10^{-5}$ in squares of size $w_b-w_1$, where $w_b=5$ and $w_1$ is
a parameter.

\begin{figure}[H]
\includegraphics[width=\linewidth]{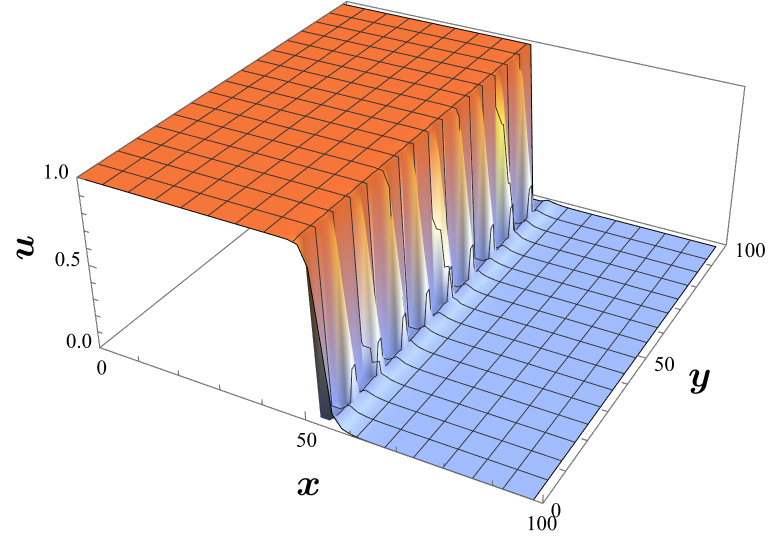}
\caption{A kink trapped by a checkerboard obstacle for $w_1 =1$
and $w_b=5$.}
\label{5w2check1}
\end{figure} 
We observed that fronts get trapped by the defect if $w_1 \le 1$
even for narrow defects such that $w_b=5$ corresponding to a single row.
An example is shown in Fig. \ref{5w2check1} which shows a plot 
of the trapped kink for $w_1 =1$ and $w_b=5$. 

\section{Conclusion}

We studied numerically the  blocking of 2D bistable reaction-diffusion
fronts by obstacles: waveguides connected to cones and checkerboard-like structures.

An analysis based on conservation law revealed the importance of the 
integral of the reaction term as a driving force of the front.
We obtained an analytical model describing the blocking and
found quantitative values for blocking of the front by a cone of
angle $\theta$ for widths smaller than $5$.
For larger waveguides the front is never blocked. 
The model illustrates the influence of geometry and nonlinearity.
These results complement the analysis of Berestycki et al. \cite{bbc16}

We expect these results to hold for general bistable nonlinearities.
We also showed that they should extend to higher dimensions in 
the sense that the front propagates only along one dimension.

This blocking of the front cannot happen for monostable reaction
terms. There, the zero state is unstable so we 
expect a secondary burst to occur when the front gets slowed down by
the junction. This secondary burst will create a new front so that
there will always be crossing, see our study \cite{ccs21}. 

\section*{Acknowledgements}
The authors benefited from the grant PAPIIT IN107624 from UNAM.
The authors thank the 
Centre R\'egional Informatique et d'Applications Num\'eriques de 
Normandie (CRIANN) for the use of its computing resources. We are also very grateful to Patrick Bousquet-Melou for adapting the code to GPU.


\section{Appendix: integrals}

From \cite{cs11} we recall the following integrals.
Noting 
$$ U(z) = {1 \over 1+\exp(z)}, ~~R(U) = U (1-U) (U-a),  $$
we have 
$$ \int_{y}^{+\infty} R[U(z)] dz = {1 \over 2 (1+ \exp(y))^2}
-{a  \over 1+ \exp(y) } ,  $$
$$ \int_{-\infty}^{+\infty} R[U(z)] dz = {1 \over 2} -a , $$
so that
$$ \int_{-\infty}^{y} R[U(z)] dz  =  {1 \over 2 } -a +
{a \over  1+ \exp(y)} -{1 \over 2 (1+ \exp(y))^2} . $$
Similarly, we have
\be
\int_{-\infty}^{+\infty} R[U(z)] z dz = -{1 \over 2},
\ee
and
\be
\begin{gathered}
\int_{y}^{+\infty} R[U(z)] z dz = {1 \over 2} [ {y \over  (1+ \exp(y))^2}\\
- {1 + 2 a y \over 1+ \exp(y)} + (2a -1) (y - \log(1+\exp(y)))] . 
\end{gathered}
\ee

\end{document}